\documentclass[aps,prd,superscriptaddress,showpacs,amsmath,amssymb]{revtex4}

\usepackage[dvips]{graphicx}
\usepackage[dvips]{color}
\usepackage{epsfig}
\usepackage{amssymb,amsmath}
\usepackage{lscape}

\newcommand{\bdk}{$B^{\pm}\to DK^{\pm}$}
\newcommand{\bdkm}{$B^{-}\to DK^{-}$}
\newcommand{\bdkp}{$B^{+}\to DK^{+}$}

\newcommand{\bdtkp}{$B^{+}\to \tilde{D}_+K^{+}$}

\newcommand{\bdskp}{$B^{+}\to D^{*}K^{+}$}

\newcommand{\bdstkp}{$B^{+}\to \tilde{D}^{*}_+K^{+}$}

\newcommand{\bdks}{$B^{\pm}\to DK^{*\pm}$}
\newcommand{\bdksm}{$B^{-}\to DK^{*-}$}
\newcommand{\bdksp}{$B^{+}\to DK^{*+}$}

\newcommand{\bdtksp}{$B^{+}\to \tilde{D}_+K^{*+}$}

\newcommand{\bddskp}{$B^{+}\to D^{(*)}K^{+}$}

\newcommand{\bddsks}{$B^{\pm}\to D^{(*)}K^{(*)\pm}$}

\newcommand{\bddsksp}{$B^{+}\to D^{(*)}K^{(*)+}$}

\newcommand{\bddstksp}{$B^{+}\to \tilde{D}^{(*)}_+K^{(*)+}$}

\newcommand{\bdksnr}{$B^{\pm}\to DK^0_S\pi^{\pm}$}
\newcommand{\bdkspnr}{$B^{+}\to DK^0_S\pi^{+}$}

\newcommand{\bdpip}{$B^{+}\to D\pi^{+}$}

\newcommand{\bdspip}{$B^{+}\to D^{*}\pi^{+}$}

\newcommand{\bndspip}{$B\to D^{*-}\pi^{+}$}

\newcommand{\bddspip}{$B^{+}\to D^{(*)}\pi^{+}$}

\newcommand{\bddstpip}{$B^{+}\to \tilde{D}^{(*)}_+\pi^{+}$}

\newcommand{\dsdpim}{$D^{*-}\to \overline{D}{}^0\pi^{-}$}
\newcommand{\dsdpims}{$D^{*-}\to \overline{D}{}^0\pi_s^{-}$}

\newcommand{\dkpp}{$\overline{D}{}^0\to K^0_S\pi^+\pi^-$}

\begin{document}

\title{Measurement of \boldmath{$\phi_3$} with Dalitz Plot
Analysis of \boldmath{\bddsksp} Decay}
\date{\today}

\affiliation{Budker Institute of Nuclear Physics, Novosibirsk}
\affiliation{Chiba University, Chiba}
\affiliation{Chonnam National University, Kwangju}
\affiliation{University of Cincinnati, Cincinnati, Ohio 45221}
\affiliation{University of Hawaii, Honolulu, Hawaii 96822}
\affiliation{High Energy Accelerator Research Organization (KEK), Tsukuba}
\affiliation{University of Illinois at Urbana-Champaign, Urbana, Illinois 61801}
\affiliation{Institute of High Energy Physics, Chinese Academy of Sciences, Beijing}
\affiliation{Institute of High Energy Physics, Vienna}
\affiliation{Institute of High Energy Physics, Protvino}
\affiliation{Institute for Theoretical and Experimental Physics, Moscow}
\affiliation{J. Stefan Institute, Ljubljana}
\affiliation{Kanagawa University, Yokohama}
\affiliation{Korea University, Seoul}
\affiliation{Kyungpook National University, Taegu}
\affiliation{Swiss Federal Institute of Technology of Lausanne, EPFL, Lausanne}
\affiliation{University of Ljubljana, Ljubljana}
\affiliation{University of Maribor, Maribor}
\affiliation{University of Melbourne, Victoria}
\affiliation{Nagoya University, Nagoya}
\affiliation{Nara Women's University, Nara}
\affiliation{National Central University, Chung-li}
\affiliation{National United University, Miao Li}
\affiliation{Department of Physics, National Taiwan University, Taipei}
\affiliation{H. Niewodniczanski Institute of Nuclear Physics, Krakow}
\affiliation{Nippon Dental University, Niigata}
\affiliation{Niigata University, Niigata}
\affiliation{Nova Gorica Polytechnic, Nova Gorica}
\affiliation{Osaka City University, Osaka}
\affiliation{Osaka University, Osaka}
\affiliation{Panjab University, Chandigarh}
\affiliation{Peking University, Beijing}
\affiliation{Princeton University, Princeton, New Jersey 08544}
\affiliation{RIKEN BNL Research Center, Upton, New York 11973}
\affiliation{Saga University, Saga}
\affiliation{University of Science and Technology of China, Hefei}
\affiliation{Seoul National University, Seoul}
\affiliation{Shinshu University, Nagano}
\affiliation{Sungkyunkwan University, Suwon}
\affiliation{University of Sydney, Sydney NSW}
\affiliation{Tata Institute of Fundamental Research, Bombay}
\affiliation{Toho University, Funabashi}
\affiliation{Tohoku Gakuin University, Tagajo}
\affiliation{Tohoku University, Sendai}
\affiliation{Department of Physics, University of Tokyo, Tokyo}
\affiliation{Tokyo Institute of Technology, Tokyo}
\affiliation{Tokyo Metropolitan University, Tokyo}
\affiliation{Tokyo University of Agriculture and Technology, Tokyo}
\affiliation{Virginia Polytechnic Institute and State University, Blacksburg, Virginia 24061}
\affiliation{Yonsei University, Seoul}
  \author{A.~Poluektov}\affiliation{Budker Institute of Nuclear Physics, Novosibirsk} 
  \author{K.~Abe}\affiliation{High Energy Accelerator Research Organization (KEK), Tsukuba} 
  \author{K.~Abe}\affiliation{Tohoku Gakuin University, Tagajo} 
  \author{I.~Adachi}\affiliation{High Energy Accelerator Research Organization (KEK), Tsukuba} 
  \author{H.~Aihara}\affiliation{Department of Physics, University of Tokyo, Tokyo} 
  \author{D.~Anipko}\affiliation{Budker Institute of Nuclear Physics, Novosibirsk} 
  \author{K.~Arinstein}\affiliation{Budker Institute of Nuclear Physics, Novosibirsk} 
  \author{T.~Aushev}\affiliation{Institute for Theoretical and Experimental Physics, Moscow} 
  \author{S.~Bahinipati}\affiliation{University of Cincinnati, Cincinnati, Ohio 45221} 
  \author{A.~M.~Bakich}\affiliation{University of Sydney, Sydney NSW} 
% ...
\collaboration{The Belle Collaboration}


\begin{abstract} 

We present a measurement of the unitarity triangle angle $\phi_3$ using a 
Dalitz plot analysis of the $K^0_S\pi^+\pi^-$ decay of the neutral $D$ meson 
from the \bddsks\ process. The method employs the interference between 
$D^0$ and $\overline{D}{}^0$ to extract the angle $\phi_3$, strong phase 
$\delta$ and the ratio $r$ of suppressed and allowed amplitudes. We apply 
this method to a 357 fb$^{-1}$ data sample collected by the Belle experiment. 
The analysis uses three modes: \bdkp, \bdskp\ with $D^{*}\to D\pi^0$, and 
\bdksp\ with $K^{*+}\to K^0_S\pi^{+}$, as well as the corresponding 
charge-conjugate modes. From a combined maximum likelihood fit to the three 
modes, we obtain $\phi_3=53^{\circ}\;^{+15^{\circ}}_{-18^{\circ}}
\mbox{(stat)}\pm 3^{\circ} \mbox{(syst)}\pm 9^{\circ}(\mbox{model})$. 
The corresponding two standard deviation interval is 
$8^{\circ}<\phi_3<111^{\circ}$. 
\end{abstract}
\pacs{12.15.Hh, 13.25.Hw, 14.40.Nd} 
\maketitle

\section{Introduction}

Determinations of the Cabibbo-Kobayashi-Maskawa
(CKM) \cite{ckm} matrix elements provide important checks on
the consistency of the standard model and ways to search
for new physics. The possibility of observing direct $CP$ violation
in $B\to D K$ decays
was first discussed by I. Bigi, A. Carter and A. Sanda \cite{bigi}.
Since then, various methods using $CP$ violation in $B\to D K$ decays have been
proposed \cite{glw,dunietz,eilam,ads} to measure the unitarity triangle
angle $\phi_3$. These methods are based on two key observations:
neutral $D^{0}$ and $\overline{D}{}^0$
mesons can decay to a common final state, and the decay
\bdkp\ can produce neutral $D$ mesons of both flavors
via $\bar{b}\to \bar{c}u\bar{s}$ and $\bar{b}\to \bar{u}c\bar{s}$ transitions,
with a relative phase $\theta_+$ between the two interfering
amplitudes that is the sum, $\delta + \phi_3$, of strong and weak interaction
phases.  For the decay \bdkm, the relative phase is
$\theta_-=\delta-\phi_3$, so both phases can be extracted
from measurements of such charge conjugate $B$ decay modes.
However, the use of branching fractions alone requires additional
information to obtain $\phi_3$.
This is provided either by determining the branching fractions of
decays to flavour eigenstates (GLW method \cite{glw})
or by using different neutral $D$ final states (ADS method \cite{ads}).

A Dalitz plot analysis of a three-body final state of the $D$ meson
allows one to obtain all the information required for determination
of $\phi_3$ in a single decay mode. The use of a Dalitz plot analysis
for the extraction of $\phi_3$ was first discussed
by D. Atwood, I. Dunietz and A. Soni, in the context of the ADS
method \cite{ads}. This technique uses the interference of
Cabibbo-favored $D^0\to K^-\pi^+\pi^0$ and doubly Cabibbo-suppressed
$\overline{D}{}^0\to K^-\pi^+\pi^0$ decays.
However, the small rate for the doubly Cabibbo-suppressed decay
limits the sensitivity of this technique.

Three body final states such as 
$K^0_S\pi^+\pi^-$ \cite{giri,binp_dalitz} have been suggested as 
promising modes for the extraction of $\phi_3$. 
In the Wolfenstein parameterization of the CKM matrix elements, 
the weak parts of the amplitudes 
that contribute to the decay \bdkp\ 
are given by $V_{cb}^*V_{us\vphantom{b}}^{\vphantom{*}}\sim A\lambda^3$ 
(for the $\overline{D}{}^0 K^+$ final state) and
$V_{ub}^*V_{cs\vphantom{b}}^{\vphantom{*}}\sim A\lambda^3(\rho+i\eta)$ (for $D^0 K^+$). 
The two amplitudes interfere as the $D^0$ and $\overline{D}{}^0$ mesons decay
into the same final state $K^0_S \pi^+ \pi^-$; 
we denote the admixed state as $\tilde{D}_+$. 
Assuming no $CP$ asymmetry in neutral $D$ decays, 
the amplitude of the $\tilde{D}_+$ decay 
as a function of Dalitz plot variables $m^2_+=m^2_{K^0_S\pi^+}$ and 
$m^2_-=m^2_{K^0_S\pi^-}$ is 
\begin{equation}
  M_+=f(m^2_+, m^2_-)+re^{i\phi_3+i\delta}f(m^2_-, m^2_+), 
\end{equation}
where $f(m^2_+, m^2_-)$ is the amplitude of the \dkpp\ decay, and
$r$ is the ratio of the magnitudes of the two interfering amplitudes. 
The value of $r$ is given by the ratio of the CKM matrix elements 
$|V_{ub}^*V_{cs\vphantom{b}}^{\vphantom{*}}|/
 |V_{cb}^*V_{us\vphantom{b}}^{\vphantom{*}}|\sim 0.38$ 
and the color suppression factor, 
and is estimated to be in the range 0.1--0.2 \cite{gronau}.

Similarly, the amplitude of the $\tilde{D}_-$ decay from \bdkm\ process is
\begin{equation}
  M_-=f(m^2_-, m^2_+)+re^{-i\phi_3+i\delta}f(m^2_+, m^2_-). 
\end{equation}
The \dkpp\ decay amplitude $f$ can be determined
from a large sample of flavor-tagged \dkpp\ decays 
produced in continuum $e^+e^-$ annihilation. Once $f$ is known, 
a simultaneous fit of $B^+$ and $B^-$ data allows the 
contributions of $r$, $\phi_3$ and $\delta$ to be separated. 
The method has a two-fold ambiguity: 
$(\phi_3,\delta)$ and $(\phi_3+180^{\circ}, \delta+180^{\circ})$
solutions cannot be separated. We always choose the solution 
with $0<\phi_3<180^{\circ}$. 
References \cite{giri} and \cite{belle_phi3_2} give  
a more detailed description of the technique. 

The method described above can be applied to other modes as well as \bdkp\ 
decay and its charge-conjugate mode (charge conjugate states are implied 
throughout the paper).
Excited states of neutral $D$ and $K$ mesons can also be used, although 
the values of $\delta$ and $r$ can differ for these decays. 
Previously the Belle~\cite{belle_phi3_2,belle_phi3_3} and 
BaBar~\cite{babar_phi3_2} collaborations
performed analyses using this technique for \bdkp\ and 
\bdskp\ decays. Both Belle~\cite{belle_phi3_bdks} and 
BaBar~\cite{babar_phi3_bdks} have also performed analyses of 
the \bdksp\ mode, but the Belle result was not combined with that 
from the \bddskp\ modes. 
The latest Belle analyses \cite{belle_phi3_3,belle_phi3_bdks} 
were based on a 253 fb$^{-1}$ data sample. In the current 
paper, we report a measurement of $\phi_3$ with the combination 
of \bdkp, \bdskp\ and \bdksp\ modes based on a 357 fb$^{-1}$ data sample. 
This analysis supersedes previous Belle results on $\phi_3$
using Dalitz plot analysis of \bddsksp\ decays. 

\section{Event selection}

We use a 357 fb$^{-1}$ data sample, corresponding to
$386\times 10^6$ $B\bar{B}$ pairs, collected by the Belle detector. 
The decay chains \bdkp, \bdskp\ with $D^*\to D\pi^0$ and \bdksp\ with 
$K^{*+}\to K^0_S\pi^+$ are selected for the analysis;
the decays \bdpip, 
\bdspip\ with $D^{*}\to D\pi^0$ 
and $\overline{B}{}^0\to D^{*+}\pi^-$ with $D^{*+}\to D\pi^+$ 
serve as control samples. The neutral $D$ meson 
is reconstructed in the $K^0_S\pi^+\pi^-$ final state in all cases. 
We also select decays of \dsdpim\ produced via the 
$e^+e^-\to c\bar{c}$ continuum process as a high-statistics 
sample to determine the \dkpp\ decay amplitude. 

The Belle detector is described in detail elsewhere \cite{belle,svd2}. 
It is a large-solid-angle magnetic spectrometer consisting of a
silicon vertex detector (SVD), a 50-layer central drift chamber (CDC) for
charged particle tracking and specific ionization measurement ($dE/dx$), 
an array of aerogel threshold \v{C}erenkov counters (ACC), time-of-flight
scintillation counters (TOF), and an array of CsI(Tl) crystals for 
electromagnetic calorimetry (ECL) located inside a superconducting solenoid coil
that provides a 1.5 T magnetic field. An iron flux return located outside 
the coil is instrumented to detect $K_L$ mesons and identify muons (KLM).

Charged tracks are required to satisfy criteria based on the 
quality of the track fit and the distance from the interaction point in both
longitudinal and transverse planes with respect to the beam axis. 
To reduce the low momentum combinatorial 
background we require each track to have a transverse momentum greater than 
100 MeV/$c$. 
Separation of kaons and pions is accomplished by combining the responses of 
the ACC and the TOF with the $dE/dx$ measurement from the CDC to 
form a likelihood $\mathcal{L}(h)$ where $h$ is a pion or a kaon. 
Charged particles are identified as pions or kaons using the likelihood ratio
$\mathcal{R}_{\rm PID}(h)=\mathcal{L}(h)/(\mathcal{L}(K)+\mathcal{L}(\pi))$.
For charged kaon identification, we require $\mathcal{R}_{\rm PID}(K)>0.7$. 
This requirement selects kaons 
with an efficiency of 80\% and pions with an efficiency of 5\%. 

Photon candidates are required to have ECL energy greater than 30 MeV. 
Neutral pion candidates are formed from pairs of photons with invariant 
masses in the range 120 to 150 MeV/$c^2$, {\it i.e.}\ less than two standard 
deviations from the $\pi^0$ mass.

Neutral kaons are reconstructed from pairs of oppositely charged tracks
without any pion PID requirement.
We require the reconstructed vertex distance from the interaction point 
in the plane transverse to the beam axis to be more than 1 mm 
and the invariant mass $M_{\pi\pi}$
to satisfy $|M_{\pi\pi}-M_{K^0_S}|<10$ MeV/$c^2$, {\it i.e.}\ 
less than four standard deviations from the nominal $K^0_S$ mass. 

\subsection{Selection of \boldmath{\dsdpim}}

To determine the \dkpp\ decay amplitude we use $D^{*\pm}$ mesons
produced via the $e^+ e^-\to c\bar{c}$ continuum process. 
The flavor of the neutral $D$ meson is tagged by the charge of the slow pion 
(which we denote as $\pi_s$) in the decay \dsdpims.

To select neutral $D$ candidates we require the invariant mass of the 
$K^0_S\pi^+\pi^-$ system to be within 9 MeV/$c^2$ of the $D^0$ mass, $M_{D^0}$.
To select events originating from a $D^{*-}$ decay 
we impose a requirement on the difference 
$\Delta M=M_{K^0_S\pi^+\pi^-\pi_s}-M_{K^0_S\pi^+\pi^-}$ of the invariant 
masses of the $D^{*-}$ and the neutral $D$ candidates: 
$144.6\mbox{ MeV}/c^2<\Delta M<146.4\mbox{ MeV}/c^2$.
The resolutions of the selection variables 
are $\sigma(\Delta M)=0.38$ MeV$/c^2$ and 
$\sigma(M_{K^0_S\pi^+\pi^-})=5.4$ MeV$/c^2$. 
The suppression of the combinatorial background from $B\bar{B}$ events
is achieved by requiring the $D^{*-}$ momentum 
in the center-of-mass (CM) frame to be greater than 2.7 GeV/$c$.

The number of events 
that pass all selection criteria is 271621. To obtain the number of background 
events in our sample we fit the $\Delta M$ distribution. The background is 
parameterized with the function 
$b(\Delta M)\sim (1/\Delta M)\sqrt{(\Delta M/m_{\pi})^2-1}$;
the function describing the signal is a combination of two Gaussian peaks with the same 
mean value. The fit finds $(261.9\pm 1.1)\times 10^3$ signal events and 
$8698\pm 77$ background events
corresponding to a background fraction of 3.2\%.

\subsection{Selection of \boldmath{\bdkp}}

The selection of $B$ candidates is based on the CM energy difference
$\Delta E = \sum E_i - E_{\rm beam}$ and the beam-constrained $B$ meson mass
$M_{\rm bc} = \sqrt{E_{\rm beam}^2 - (\sum p_i)^2}$, where $E_{\rm beam}$ 
is the CM beam 
energy, and $E_i$ and $p_i$ are the CM energies and momenta of the
$B$ candidate decay products. We select events with $M_{\rm bc}>5.2$~GeV/$c^2$
and $|\Delta E|<0.2$~GeV for the analysis. 
In addition, we impose a requirement on the invariant mass of the 
neutral $D$ candidate: 
$|M_{K^0_S\pi^+\pi^-}-M_{D^0}|<11$~MeV/$c^2$. 


To suppress background from $e^+e^-\to q\bar{q}$ ($q=u, d, s, c$) 
continuum events, we require $|\cos\theta_{\rm thr}|<0.8$, 
where $\theta_{\rm thr}$ is the angle between the thrust axis of 
the $B$ candidate daughters and that of the rest of the event. 
For additional background rejection, we 
use a Fisher discriminant composed of 11 parameters \cite{fisher}: 
the production angle of the $B$ candidate, the angle of the $B$ thrust 
axis relative to the beam axis and nine parameters representing 
the momentum flow in the event relative to the $B$ thrust axis in the CM frame.
We apply a requirement on the Fisher 
discriminant that retains 90\% of the signal and rejects 40\% of the 
remaining continuum background. 

The $\Delta E$ and $M_{\rm bc}$ distributions for \bdkp\ candidates are
shown in Fig.~\ref{signal_mbcde}a,b. The peak in the $\Delta E$ distribution at 
$\Delta E=50$~MeV is due to \bdpip\ decays where the pion is misidentified
as a kaon.
The \bdkp\ selection efficiency (11\%) is determined from 
a Monte Carlo (MC) simulation. The number of events in the signal 
box ($|\Delta E|<30$~MeV, $5.27$~GeV/$c^2<M_{\rm bc}<5.3$~GeV/$c^2$) is 470. 

For the selected events a two-dimensional unbinned
maximum likelihood fit in the variables $M_{\rm bc}$ and $\Delta E$ 
is performed. The resulting signal and background density functions 
are used in the Dalitz plot fit to 
obtain the event-by-event signal to background ratio. 
To parametrize the shape of the $\Delta E-M_{\rm bc}$ distribution, 
we use two-dimensional Gaussian peaks for the signal contribution 
and \bdpip\ background. 
The non-peaking background is parametrized by the sum of two 
components: the product of an empirical shape introduced by ARGUS \cite{argus} 
as a function of $M_{\rm bc}$ and a linear function in $\Delta E$, and 
the product of a Gaussian distribution in $M_{\rm bc}$ and a 
linear function in $\Delta E$. 
Only the region with $\Delta E>-0.1$~GeV is used in the fit; the region 
with $\Delta E<-0.1$~GeV includes a large fraction of events from $B\to D^*K$
decay with a lost pion that do not contribute to the signal box. 
The number of events in the signal peak obtained from the fit is 
$331\pm 23$; the event purity in the signal box is 67\%. 

\subsection{Selection of \boldmath{\bdskp}}

For the selection of \bdskp\ events, in addition to the 
requirements described above, we require that the mass difference 
$\Delta M=M_{K^0_S\pi^+\pi^-\pi^0}-M_{K^0_S\pi^+\pi^-}$ of 
neutral $D^{*}$ and $D$ candidates satisfies 
$140\mbox{ MeV}/c^2<\Delta M<145\mbox{ MeV}/c^2$.
Figures \ref{signal_mbcde}c,d show the $\Delta E$ and $M_{\rm bc}$
distributions for \bdskp\ candidates. The selection 
efficiency is 6.2\%. The number of events in the signal box is
111. The parametrization of background and signal shapes is similar 
to that in the \bdkp\ case. The number of events in the signal 
peak obtained from the fit is $81\pm 11$; the event purity in the 
signal box is 77\%. 

\subsection{Selection of \boldmath{\bdksp}}

\begin{figure}[p]
  \epsfig{figure=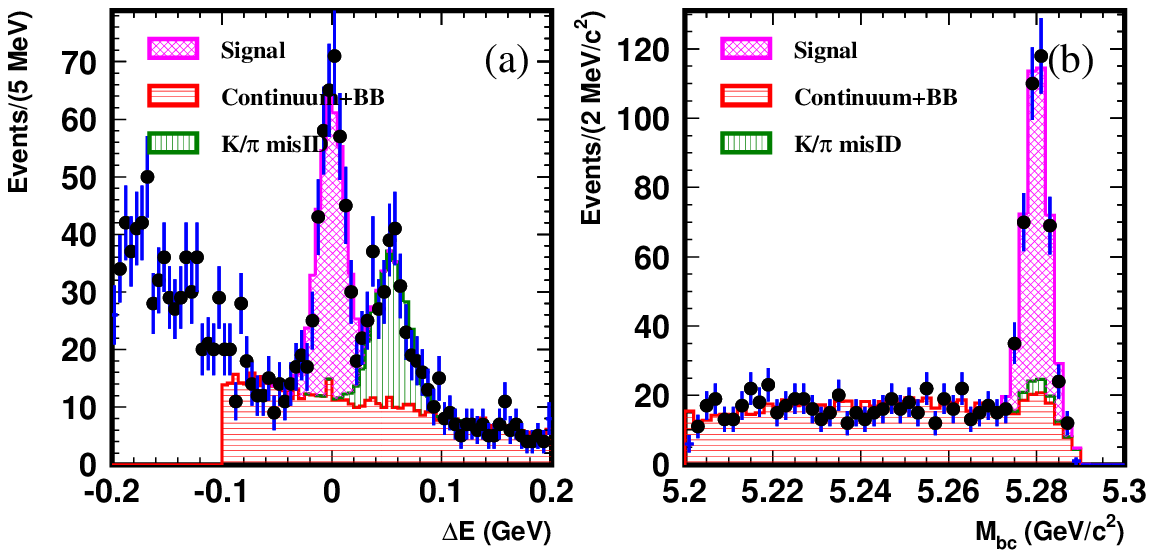,width=0.8\textwidth}
  \epsfig{figure=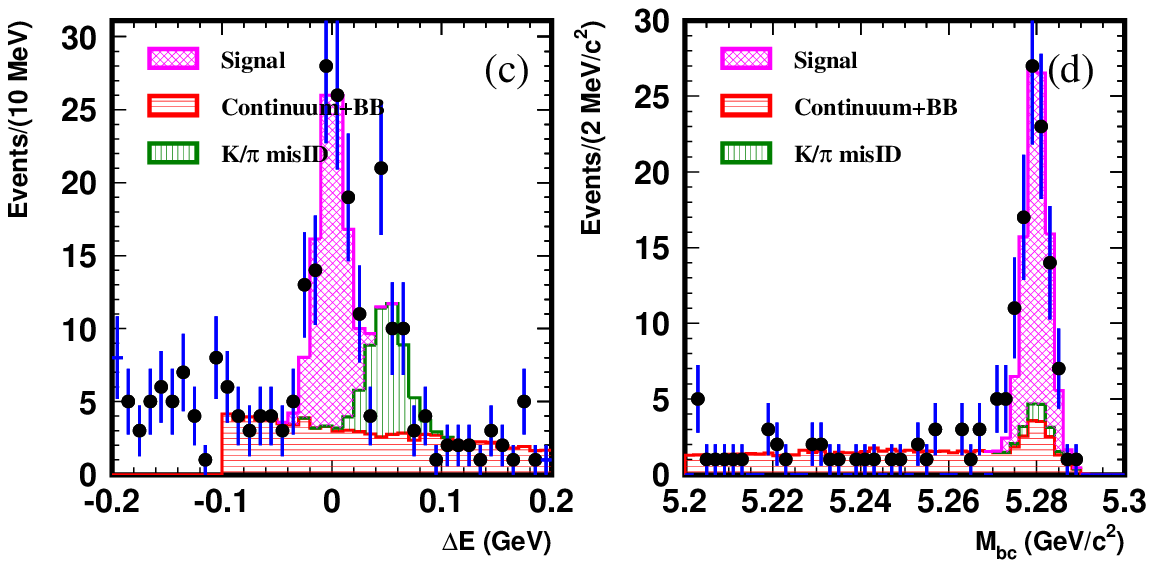,width=0.8\textwidth}
  \epsfig{figure=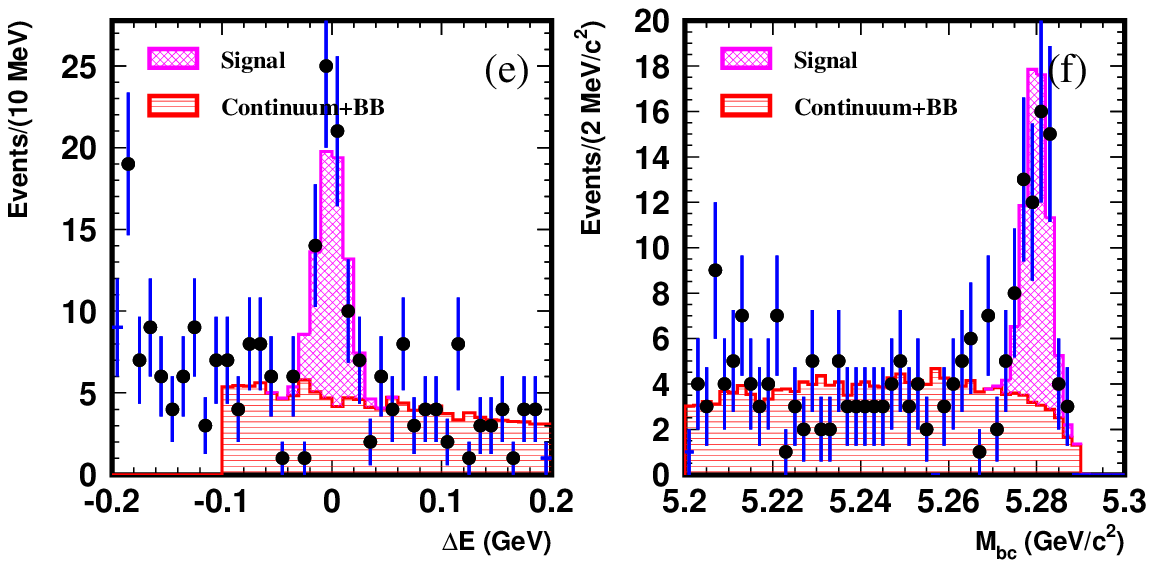,width=0.8\textwidth}
  \caption{$\Delta E$ and $M_{\rm bc}$ distributions for the 
           (a,b) \bdkp, (c,d) \bdskp, and (e,f) \bdksp\ event samples. 
           Points with error bars are the data, and
           the histogram is the result of a MC simulation according to the 
           fit result. }
  \label{signal_mbcde}
\end{figure}

For \bdksp\ decay candidates, in addition to the requirements placed on the
\bdkp\ mode, $K^{*+}$ candidate selection is performed. 
To select $K^{*+}$ meson candidates, we require the $K^0_S\pi^{+}$ 
invariant mass to be within 
50 MeV/$c^2$ of the $K^{*+}$ nominal mass. For continuum background 
suppression, in addition to the $|\cos\theta_{\rm thr}|<0.8$
condition, we require that the $K^{*+}$ helicity angle satisfy 
$|\cos\theta_{\rm hel}|>0.4$. The $K^{*+}$ helicity angle is defined
as the angle between the axis of the $K^{*+}$ decay products, and the
momentum of the $B$ meson, in the $K^{*+}$ rest frame. 
We also apply a requirement on the Fisher 
discriminant that retains 95\% of the signal and rejects 30\% of the 
remaining continuum background. 
Figures \ref{signal_mbcde}e,f show the $\Delta E$ and $M_{\rm bc}$
distributions for \bdksp\ candidates. The selection 
efficiency is 4.1\%. The number of events in the signal box is
78. The parametrization of background and signal shapes is similar 
to that in the \bdkp\ case, but without the contribution of $K/\pi$ misidentification
background. The number of events in the signal 
peak obtained from the fit is $54\pm 8$; the event purity in the 
signal box is 65\%.

\section{Determination of the \boldmath{\dkpp} decay amplitude}

\label{section_d0_fit}

The amplitude $f$ for the \dkpp\ decay is described
by a coherent sum of two-body decay amplitudes and one non-resonant 
decay amplitude,
\begin{equation}
  f(m^2_+, m^2_-) = \sum\limits_{j=1}^{N} a_j e^{i\alpha_j}
  \mathcal{A}_j(m^2_+, m^2_-)+
    b e^{i\beta}, 
  \label{d0_model}
\end{equation}
where $N$ is the total number of resonances, 
$\mathcal{A}_j(m^2_+, m^2_-)$ is the matrix element, $a_j$ and 
$\alpha_j$ are the amplitude and phase of the matrix element, respectively, 
of the $j$-th resonance, and $b$ and $\beta$ are the amplitude
and phase of the non-resonant component. The total phase and amplitude 
are arbitrary. To be consistent with other analyses 
\cite{babar_phi3_2,dkpp_cleo}
we have chosen the $\overline{D}{}^0\to K^0_S\rho$ 
mode to have unit amplitude and zero relative phase. 
The description of the matrix elements follows Ref.~\cite{cleo_model}. 

We use a set of 18 two-body amplitudes. 
These include five Cabibbo-allowed amplitudes: $K^*(892)^+\pi^-$, 
$K^*(1410)^+\pi^-$, 
$K_0^*(1430)^+\pi^-$, 
$K_2^*(1430)^+\pi^-$ and $K^*(1680)^+\pi^-$;  
their doubly Cabibbo-suppressed partners; and eight amplitudes with
$K^0_S$ and a $\pi\pi$ resonance:
$K^0_S\rho$, $K^0_S\omega$, $K^0_Sf_0(980)$, $K^0_Sf_2(1270)$, 
$K^0_Sf_0(1370)$, $K^0_S\rho(1450)$, $K^0_S\sigma_1$ and $K^0_S\sigma_2$. 
The differences from our previous publication \cite{belle_phi3_2} are, 
i) the addition of  $K^0_S\rho(1450)$, $K^*(1410)^+\pi^-$ and the 
   corresponding doubly Cabibbo-suppressed mode, 
ii) the use of the Gounaris-Sakurai \cite{gounaris} amplitude description for 
the $K^0_S\rho$ and $K^0_S\rho(1450)$ contributions, and 
iii) the mass and width for the $f_0(1370)$ state are now 
taken from Ref.~\cite{aitala} 
($M=1434$ MeV$/c^2$, $\Gamma=173$ MeV$/c^2$).

We use an unbinned maximum likelihood technique to fit the Dalitz plot 
distribution to the model described by Eq.~\ref{d0_model}. 
We minimize the negative 
logarithm of the likelihood function in the form
\begin{equation}
  -2 \log L = -2\left[\sum\limits^n_{i=1}\log p(m^2_{+,i}, m^2_{-,i}) - 
  \log\int\limits_D p(m^2_+, m^2_-)\; dm^2_+\; d m^2_-\right], 
  \label{log_l_2d}
\end{equation}
where $i$ runs over all selected event candidates, and
$m^2_{+,i}$, $m^2_{-,i}$ are measured Dalitz plot
variables. The integral in the second term accounts for the overall 
normalization of the probability density. 

The Dalitz plot density is represented by
\begin{equation}
  p(m^2_+, m^2_-) = \varepsilon(m^2_+, m^2_-)
  |f(m^2_+, m^2_-)|^2+B(m^2_+, m^2_-),
  \label{density}
\end{equation}
where $f(m^2_+, m^2_-)$ is the decay amplitude described 
by Eq.~\ref{d0_model}, $\varepsilon(m^2_+, m^2_-)$ is the efficiency, and
$B(m^2_+, m^2_-)$ is the background density. 
To take into account the finite momentum resolution of the detector, 
the Dalitz plot density described by Eq.~\ref{density} is convolved 
with a resolution function. 
The free parameters of the minimization are the amplitudes
$a_j$ and phases $\alpha_j$ of the resonances (except for the $K^0_S\rho$
component, for which the parameters are fixed), 
the amplitude $b$ and phase $\beta$ of the non-resonant component
and the masses and widths of the $\sigma_1$ and $\sigma_2$ scalars. 

\begin{figure}
  \epsfig{figure=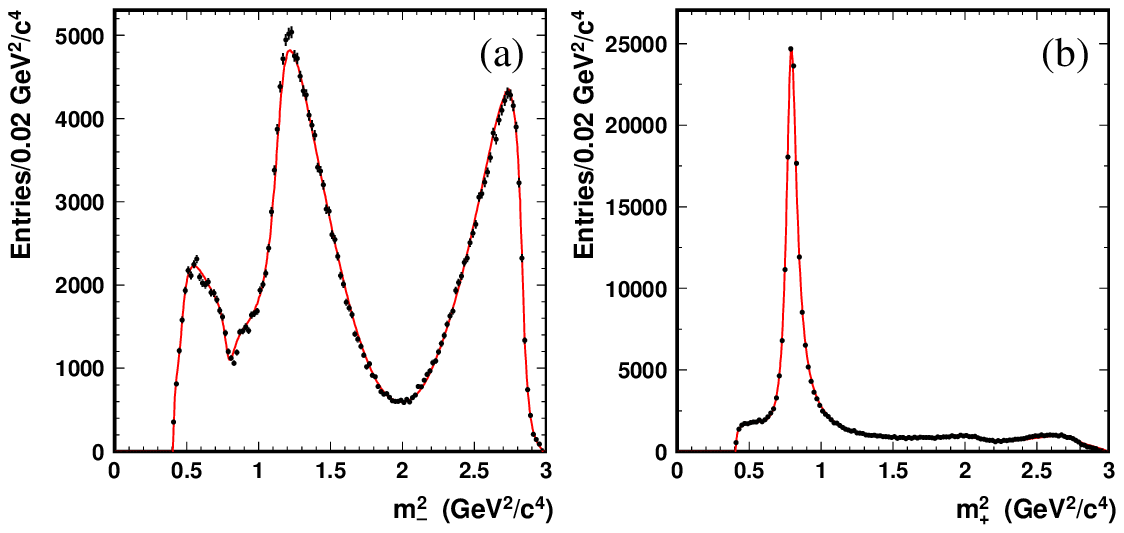,width=0.8\textwidth}
  \epsfig{figure=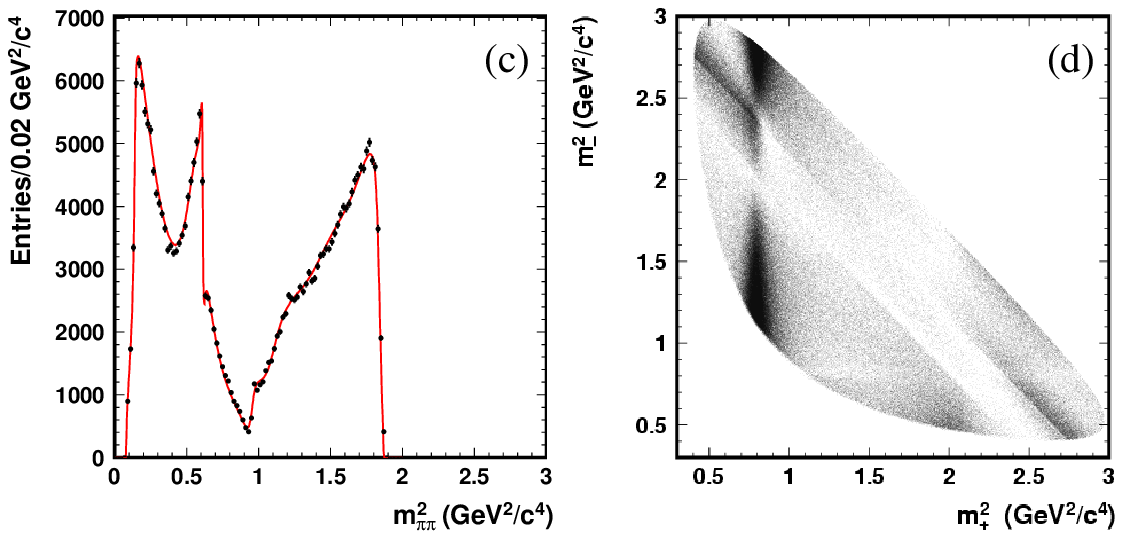,width=0.8\textwidth}
  \caption{(a) $m_-$, (b) $m_+$, (c) $m_{\pi\pi}$ and (d)
           Dalitz plot distribution for \dsdpims, \dkpp\ 
           decays from the $e^+e^-\to c\bar{c}$ continuum process. 
	   The points with error bars show the data;
           the smooth curve is the fit result.}
  \label{ds2dpi_plot}
\end{figure}

The procedures for determining the background
density, the efficiency, and the resolution of the squared invariant
mass, are the same as in the previous analyses 
\cite{belle_phi3_2,belle_phi3_3}. 
The \dkpp\ Dalitz plot distribution, as well as
its projections with the fit results superimposed, are shown in 
Fig.~\ref{ds2dpi_plot}. 
The fit results are given in Table~\ref{dkpp_table}. 
The fit fractions quoted in Table~\ref{dkpp_table} for specific
modes are defined as the integral of the absolute value squared 
of the individual mode divided by the integral of the absolute value squared
of the total amplitude. Due to the interference effects these 
fit fractions may not sum to unity. 

The parameters obtained for the $\sigma_1$ resonance 
($M_{\sigma_1}=519\pm 6$ MeV/$c^2$, $\Gamma_{\sigma_1}=454\pm 12$ MeV/$c^2$) 
are similar to those observed by other experiments~\cite{dkpp_cleo, aitala2}.
The second scalar term $\sigma_2$ is introduced to account for
structure observed at $m^2_{\pi\pi} \sim 1.1\,\mathrm{GeV}^2/c^4$:
the fit finds a small but significant contribution with
$M_{\sigma_2}=1050\pm 8$ MeV/$c^2$, $\Gamma_{\sigma_2}=101\pm 7$ MeV/$c^2$
(the errors are statistical only).
The large peak in the $m^2_+$ distribution 
corresponds to the dominant $\overline{D}{}^0\to K^*(892)^+\pi^-$ mode. 
The minimum in the $m^2_-$ distribution at 0.8~GeV$^2/c^4$
is due to destructive interference with the doubly Cabibbo 
suppressed $\overline{D}{}^0\to K^*(892)^-\pi^+$ amplitude. In the $m^2_{\pi\pi}$
distribution, the $\overline{D}{}^0\to K^0_S\rho$ contribution 
is visible around 0.5~GeV$^2/c^4$
with a steep edge on the upper side due to interference with 
$\overline{D}{}^0\to K^0_S\omega$. The minimum around 0.9~GeV$^2/c^4$ is due to 
the decay $\overline{D}{}^0\to K^0_S f_0(980)$ interfering destructively with
other modes.

We perform a $\chi^2$ test to check the quality of the 
fit, dividing the Dalitz plot into square regions 
$0.05\times 0.05$ GeV$^2/c^4$. The test finds a reduced chi-square 
$\chi^2/ndf=2.72$ for 1081 degrees of freedom ($ndf$), which is large. 
Examining Fig.~\ref{ds2dpi_plot}, we find that the main features of the 
Dalitz plot are well-reproduced, with some significant but numerically
small discrepancies at peaks and dips of the distribution. 
In our final results we include a conservative contribution to the 
systematic error due to uncertainties in the $\overline{D}{}^0$ decay model, 
discussed in Section~\ref{section_model}. 
 

\begin{table}
\caption{Fit results for \dkpp\ decay. Errors are statistical only.}
\label{dkpp_table}
\begin{tabular}{|l|c|c|c|} \hline
Intermediate state           & Amplitude 
			     & Phase ($^{\circ}$) 
			     & Fit fraction
			     \\ \hline

$K^0_S \sigma_1$               & $1.43\pm 0.07$
			     & $212\pm 3$
			     & 9.8\%
                             \\

$K^0_S\rho^0$                  & $1.0$ (fixed)                                 
                             & 0 (fixed)
			     & 21.6\%
                             \\

$K^0_S\omega$                  & $0.0314\pm 0.0008$
			     & $110.8\pm 1.6$
			     & 0.4\%
                             \\

$K^0_S f_0(980)$               & $0.365\pm 0.006$
			     & $201.9\pm 1.9$
			     & 4.9\%
                             \\

$K^0_S \sigma_2$               & $0.23\pm 0.02$
			     & $237\pm 11$
			     & 0.6\%
                             \\

$K^0_S f_2(1270)$              & $1.32\pm 0.04$
			     & $348\pm 2$
			     & 1.5\%
                             \\

$K^0_S f_0(1370)$              & $1.44\pm 0.10$
			     & $82\pm 6$
			     & 1.1\%
                             \\

$K^0_S \rho^0(1450)$           & $0.66\pm 0.07$
			     & $9\pm 8$
			     & 0.4\%
                             \\

$K^*(892)^+\pi^-$            & $1.644\pm 0.010$
			     & $132.1\pm 0.5$
			     & 61.2\%
                             \\ 

$K^*(892)^-\pi^+$            & $0.144\pm 0.004$
			     & $320.3\pm 1.5$
			     & 0.55\%
                             \\

$K^*(1410)^+\pi^-$	     & $0.61\pm 0.06$
			     & $113\pm 4$
			     & 0.05\%
			     \\

$K^*(1410)^-\pi^+$	     & $0.45\pm 0.04$
			     & $254\pm 5$
			     & 0.14\%
			     \\

$K_0^*(1430)^+\pi^-$         & $2.15\pm 0.04$
			     & $353.6\pm 1.2$
			     & 7.4\%
                             \\

$K_0^*(1430)^-\pi^+$         & $0.47\pm 0.04$
			     & $88\pm 4$
			     & 0.43\%
                             \\

$K_2^*(1430)^+\pi^-$         & $0.88\pm 0.03$
			     & $318.7\pm 1.9$
			     & 2.2\%
                             \\

$K_2^*(1430)^-\pi^+$         & $0.25\pm 0.02$
			     & $265\pm 6$
			     & 0.09\%
                             \\

$K^*(1680)^+\pi^-$           & $1.39\pm 0.27$
			     & $103\pm 12$
			     & 0.36\%
                             \\

$K^*(1680)^-\pi^+$           & $1.2\pm 0.2$
			     & $118\pm 11$
			     & 0.11\%
                             \\

non-resonant                 & $3.0\pm 0.3$
			     & $164\pm 5$
			     & 9.7\%
                             \\ 
\hline
\end{tabular}
\end{table}

\section{Dalitz plot analysis of \boldmath{\bddsksp} decays}

\label{dalitz_analysis}


In our previous analyses, the two Dalitz distributions corresponding 
to the decays of $B^+$ and $B^-$ were fitted simultaneously to give the
parameters $r$, $\phi_3$ and $\delta$. Confidence intervals were then 
calculated using a frequentist technique, relying on toy MC simulation. 
In this approach, there was a bias in the fitted value of the 
(positive definite) parameter $r$, and the errors on $\phi_3$ and 
$\delta$ were also $r$-dependent. 

In the present analysis, we use a method similar to that of BaBar
\cite{babar_phi3_2}: fitting the Dalitz distributions of the 
$B^+$ and $B^-$ samples separately, using Cartesian parameters 
$x_{\pm}=r_{\pm}\cos(\pm\phi_3+\delta)$ and 
$y_{\pm}=r_{\pm}\sin(\pm\phi_3+\delta)$, where the indices ``$+$" and 
``$-$" correspond to $B^+$ and $B^-$ decays, respectively. 
Note that in this approach the amplitude ratios ($r_+$ and $r_-$) are 
not constrained to be equal for the $B^+$ and $B^-$ samples. 
Confidence intervals in $r$, $\phi_3$ and $\delta$ are then obtained 
from the $(x_{\pm},y_{\pm})$ using a frequentist technique. The advantage
of this approach is low bias and simple distributions of the fitted 
parameters, at the price of fitting in a space with higher dimensionality 
$(x_+,y_+,x_-,y_-)$ than that of the physical parameters 
$(r, \phi_3, \delta)$; see Section~\ref{section_stat}.

The fit to a single Dalitz distribution with free parameters 
$x$ and $y$ is performed by minimizing the negative unbinned likelihood 
function 
\begin{equation}
  -2 \log L = -2\left[\sum\limits^n_{i=1}
  \log p(m^2_{+,i}, m^2_{-,i}, \Delta E_i, M_{{\rm bc},i}) - 
  \log\int\limits_D p(m^2_+, m^2_-, \Delta E, M_{\rm bc}))\;
  dm^2_+\; d m^2_-\; d\Delta E\; dM_{\rm bc}
  \right], 
  \label{log_l_4d}
\end{equation}
with the Dalitz distribution density $p$ represented as 
\begin{equation}
  p(m^2_+, m^2_-, \Delta E, M_{\rm bc}) = \varepsilon(m^2_+, m^2_-)
  |f(m^2_+, m^2_-) + (x+iy)f(m^2_-, m^2_+)|^2 F_{\rm sig}(\Delta E, M_{\rm bc}) + 
  F_{\rm bck}(m^2_+, m^2_-, \Delta E, M_{\rm bc}), 
  \label{dalitz_density}
\end{equation}
where $F_{\rm sig}$ is the signal distribution as a function of $\Delta E$ and 
$M_{\rm bc}$ (represented by the product of two Gaussian shapes), 
$F_{\rm bck}$ is the distribution of the background, 
and $\varepsilon(m^2_+, m^2_-)$
is the efficiency distribution over the phase space. 
As in the study of the sample from continuum \dsdpims\ decays, 
the finite momentum resolution is taken into account by convolving the 
function (\ref{dalitz_density}) with a Gaussian resolution function. 
The efficiency and the momentum resolution were extracted from the signal 
MC sample, where the neutral $D$ meson decays according to phase space. 
The determination of the background contribution and efficiency 
profile is described below. 

\subsection{Backgrounds}

To take backgrounds into account in the analysis, their Dalitz plot 
distributions, $\Delta E-M_{\rm bc}$ distributions (which in general may
depend on Dalitz plot region) and relative fractions have to be known. 
The backgrounds are divided into three categories:
\begin{itemize}
  \item Continuum background
  \item \bddspip\ background (with $\pi/K$ misidentification). This background 
       is relevant only for the \bdkp\ and \bdskp\ modes. 
  \item Other $b\bar{b}$ backgrounds. 
\end{itemize}

Continuum background (from the process $e^+e^-\to q\bar{q}$, where 
$q=u, d, s, c$) gives the largest contribution. 
It includes both pure  
combinatorial background, and continuum $D^0$ mesons combined with a 
random kaon. This type of background is studied in data with
$\cos\theta_{\rm thr}$ and Fisher discriminant requirements
applied to select continuum events. To check that the Dalitz plot shape 
of the continuum background selected 
using these requirements corresponds to that of the signal region, 
we use a MC sample that includes $e^+e^-\to q\bar{q}$ 
($q=u,d,s,c$) decays. The Dalitz plot distribution of the continuum background
is parametrized by a third-order polynomial in the variables 
$m^2_+$ and $m^2_-$ (which represents the combinatorial component)  
and a sum of $D^0$ and $\overline{D}{}^0$ shapes for real neutral $D$ mesons 
combined with random kaons.

The \bddspip\ process with a pion misidentified as a kaon is suppressed by the
requirements on the $K/\pi$ identification variable $\mathcal{R}_{\rm PID}$ 
and the CM energy difference. The fraction of this background is obtained 
by fitting the $\Delta E-M_{\rm bc}$ 
distribution; the corresponding Dalitz plot distribution is that of a 
$D^0$ without the opposite flavor admixture. 

Other $B\bar{B}$ backgrounds, of which the dominant fraction
comes from the decay of $D^{(*)0}$ from one $B$ meson, 
with some particles taken from the other $B$ decay, 
are investigated with $e^+e^-\to\Upsilon(4S)\to B\bar{B}$ MC events. 
The Dalitz plot distribution 
of the $b\bar{b}$ background is parameterized by a second-order polynomial 
(for the \bdkp\ and \bdksp\ modes) or by a linear function 
(for the \bdskp\ mode)
of $m^2_+$ and $m^2_-$, plus a ``correct-flavor" $D^0$
shape ($\overline{D}{}^0$ for $B^+$ data and $D^0$ for $B^-$ data). 

\subsection{Efficiency}

Knowledge of the absolute value of the reconstruction efficiency is not essential
for our analysis. However, the relative variations of the efficiency over
the $D^0$ decay phase space can affect the fit result. The shapes of the 
efficiency across the phase space are studied for each mode using signal 
MC samples with a constant \dkpp\ decay amplitude. The efficiency profiles are 
fitted with third-order polynomial functions of $m_+^2$ and $m_-^2$ symmetric 
under exchange of $\pi^+$ and $\pi^-$. The efficiency is nearly flat over
the phase space, falling by 10--20\% at its edges (relative to the efficiency 
at the center). 

\subsection{Control sample fits}

\label{section_control}

To test the consistency of the fitting procedure, the same procedure was 
applied to the \bddstpip\ and \bndspip\ control 
samples 
as to the \bddstksp\ signal. 
For decays to which only one $D$ flavor can contribute, the fit 
should return values of the amplitude ratio $r$ consistent with zero. 
In the case of \bddstpip\, a small amplitude ratio $r\sim 0.01$ is expected 
(due to the small ratio of the weak coefficients 
$|V_{ub}^{\vphantom{*}} V^*_{cd}|/
 |V_{cb}^{\vphantom{*}} V^*_{ud}|\sim 0.02$ and the additional color 
suppression factor as in the case of \bdtkp). Deviations 
from these values can appear if the Dalitz plot distribution is not 
well described by the fit model. 

For the control sample fits, we treat $B^+$ and $B^-$ data separately, 
to check for the absence of $CP$ violation. 
The free parameters of the Dalitz plot fit are $x_{\pm}$ and 
$y_{\pm}$. 

\begin{table}
  \caption{Results of fits to test samples in parameters $(x,y)$. Errors 
are statistical only.}
  \label{test_fit_table}
  \begin{tabular}{|l|c|c|c|c|}\hline
    Mode    & $x_-$ & $y_-$ & $x_+$ & $y_+$ \\ \hline
    \bdpip   & $-0.030\pm 0.015$ & $-0.014\pm 0.016$ & 
              $-0.003\pm 0.014$ & $-0.021\pm 0.018$ \\ 
    \bdspip  & $-0.007\pm 0.035$   & $-0.078\pm 0.039$  & 
              $\phantom{-}0.004\pm 0.032$   & $\phantom{-}0.022\pm 0.039$  \\ 
    \bndspip & $\phantom{-}0.016\pm 0.030$   & $\phantom{-}0.014\pm 0.030$  & 
              $\phantom{-}0.026\pm 0.026$   & $\phantom{-}0.009\pm 0.036$  \\ \hline
  \end{tabular}
\end{table}

\begin{figure}
  \epsfig{figure=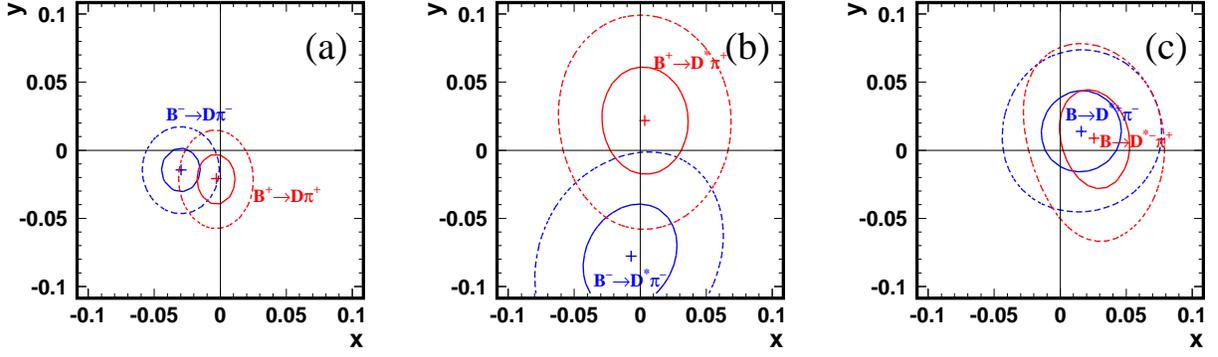,width=0.9\textwidth}
  \caption{Results of test sample fits with free parameters 
           $x=r\cos\theta$ and $y=r\sin\theta$ for (a) \bdpip, 
           (b) \bdspip\ and (c) \bndspip\ samples, separately for $B^-$
           and $B^+$ data. Contours indicate integer multiples of the 
           standard deviation. }
  \label{test_fit}
\end{figure}

The fit results for the three test samples are presented in 
Table~\ref{test_fit_table}; contour plots showing integer 
multiples of the standard deviation in the $x_{\pm}$ and $y_{\pm}$ 
variables for the three test samples are shown in Fig.~\ref{test_fit}. 
The results are consistent with $r\sim 0.01$ for the \bdpip\ and \bdspip\ modes
and with zero for the \bndspip\ mode. 

\subsection{Signal fit results}

\begin{figure}[p]
  \epsfig{figure=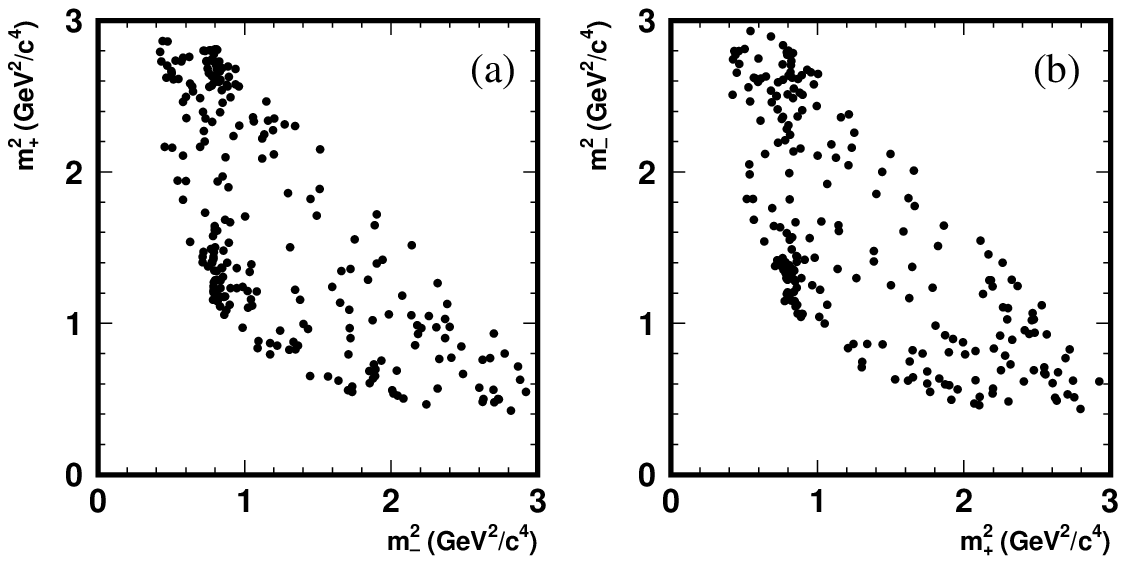,width=0.8\textwidth}
  \epsfig{figure=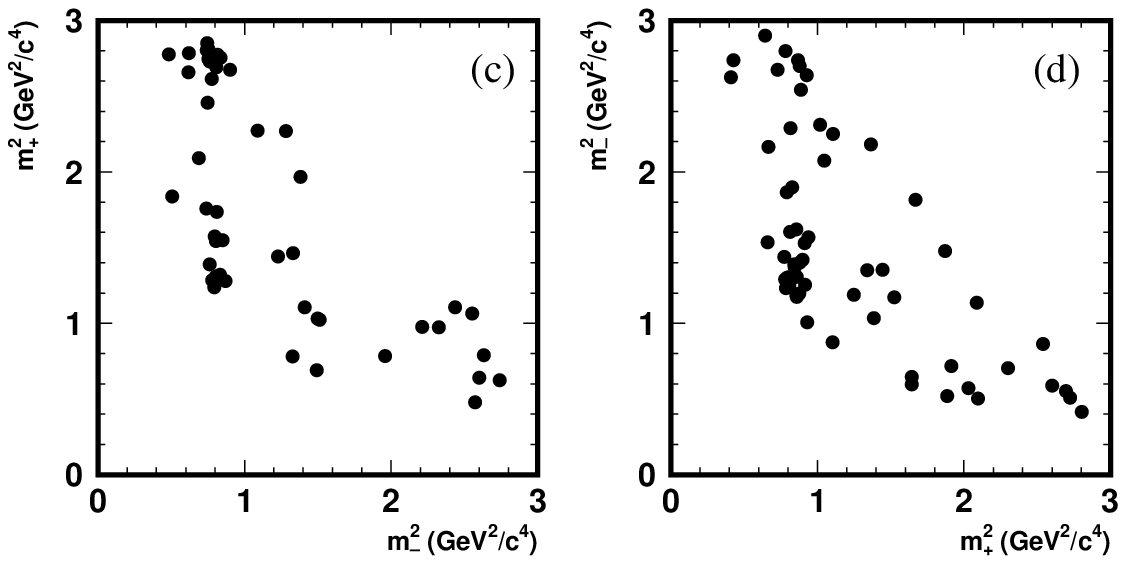,width=0.8\textwidth}
  \epsfig{figure=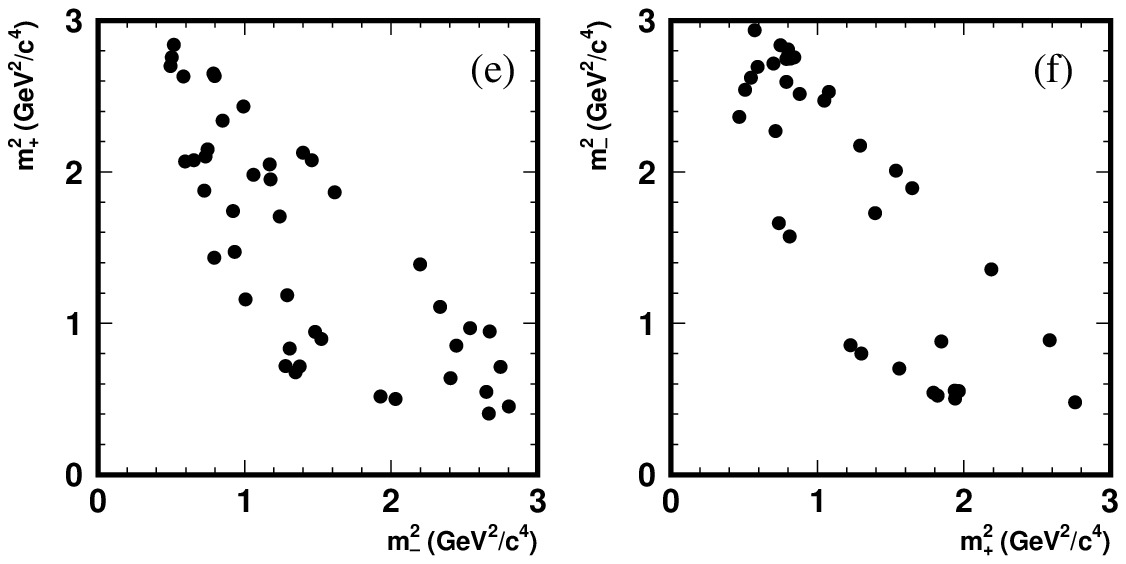,width=0.8\textwidth}
  \caption{Dalitz plots for the neutral $D$ meson from $B^-$ and 
           $B^+$ decays, respectively, for the (a,b) \bdkp, 
           (c,d) \bdskp, and (e,f) \bdksp\ modes. 
           Note that the axes are flipped in the case of $B^-$
           with respect to $B^+$. 
           }
  \label{dalitz}
\end{figure}

\begin{table}
  \caption{Results of the signal fits in parameters $(x,y)$. Errors are 
statistical only. }
  \label{sig_fit_table}
  \begin{tabular}{|l||c|c|c|c|}\hline
    Mode    & $x_-$ & $y_-$ & $x_+$ & $y_+$ \\ \hline
    \bdkp    & $\phantom{-}0.025^{+0.072}_{-0.080}$  & $\phantom{-}0.170^{+0.093}_{-0.117}$ & 
              $-0.135^{+0.069}_{-0.070}$ & $-0.085^{+0.090}_{-0.086}$ \\ 
    \bdskp   & $-0.128^{+0.167}_{-0.146}$ & $-0.339^{+0.172}_{-0.158}$ & 
              $\phantom{-}0.032^{+0.120}_{-0.116}$  & $\phantom{-}0.008^{+0.137}_{-0.136}$ \\ 
    \bdksp   & $-0.784^{+0.249}_{-0.295}$ & $-0.281^{+0.440}_{-0.335}$ & 
              $-0.105^{+0.177}_{-0.167}$ & $-0.004^{+0.164}_{-0.156}$\\ \hline
  \end{tabular}
\end{table}

\begin{figure}
  \epsfig{figure=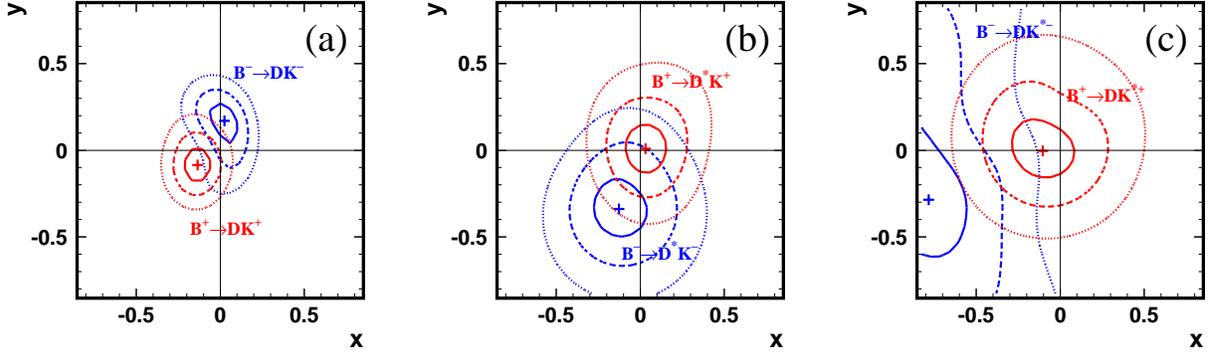,width=0.9\textwidth}
  \caption{Results of signal fits with free parameters 
           $x=r\cos\theta$ and $y=r\sin\theta$ for (a) \bdkp, 
           (b) \bdskp\ and (c) \bdksp\ samples, separately for $B^-$
           and $B^+$ data. Contours indicate integer multiples of the 
           standard deviation. }
  \label{sig_fit}
\end{figure}

The Dalitz distributions of \bdtkp, \bdstkp, and \bdtksp\ modes for the events 
in the signal region used in the fits are shown in Fig.~\ref{dalitz}.
The results of the separate $B^+$ and $B^-$ data fits are shown in 
Fig.~\ref{sig_fit}. The plots show the constraints on 
the parameters $x$ and $y$ for the \bdtkp, \bdstkp\ and \bdtksp\ samples. 
The values of the fit parameters $x_{\pm}$ and $y_{\pm}$ are 
listed in Table~\ref{sig_fit_table}. 

The fit to the \bdkp\ sample yields $r_+$ and $r_-$ values close to 
each other ($r_+=0.172$ and $r_-=0.160$) with significantly different values
of the total phase $\theta_+$ and $\theta_-$, which is an indication
of $CP$ violation. For the \bdskp\ and \bdksp\ samples, 
the values of $r_+$ and $r_-$ differ (with large errors); in both cases, 
$r_+$ is within a standard deviation of
zero, so the corresponding phase $\theta_+$ is poorly determined and the 
significance of $CP$ violation is low. 

\subsection{Evaluation of statistical error}

\label{section_stat}

We use a frequentist technique to evaluate the 
statistical significance of the measurements. This method requires 
knowledge of the probability density function (PDF) of the 
reconstructed parameters $x$ and $y$ as a function of the true parameters
$\bar{x}$ and $\bar{y}$. 
To obtain this PDF, we employ a ``toy" MC technique that uses a
simplified MC simulation of the experiment which incorporates
the same efficiencies, resolution and backgrounds as
used in the fit to the experimental data. 

For each mode, 4000 toy MC experiments are made with generated 
$(\bar{x},\bar{y})$ equal to the experimental fit results. 
The PDFs obtained are approximately Gaussian, but we observe the width
of the distribution to vary with $\bar{x}$ and $\bar{y}$. 
There is also an offset in the parameters $x,y$ introduced by the fit 
procedure. The correlations between $x$ and $y$ are negligible. 
For the \bdkp\ and \bdskp\ modes, 
the deviations from the Gaussian shape are small, and taking into 
account the tails of the distributions has only a minor effect on the 
fit result. However, the result of the measurement in the \bdksm\ mode
has a relatively large value of $r$ ($r_-=0.83\pm 0.28$). For such a large 
value, the toy MC study shows the existence of significant tails in 
the observed $x$ and $y$ distributions. 
To parametrize the shapes of the $x$ and $y$ distributions 
for the \bdksp\ sample we perform additional toy MC simulation for
different generated $\bar{x}$ and $\bar{y}$, and fit the distributions 
with the sum of two Gaussians (``narrow" and ``wide"). 
To obtain the PDF shape 
for any given set of $(\bar{x},\bar{y})$, linear interpolation in 
$\bar{x}$ and $\bar{y}$ (for the bias of the wide Gaussian relative to the 
narrow one) and in $\bar{x}^2$ and $\bar{y}^2$ 
(for the width and fraction of the wide Gaussian) are used. 

To obtain the $(\bar{x},\bar{y})$ dependence of $\sigma_x$ and $\sigma_y$
and the offset in $x$ and $y$, we generate toy MC experiments at different 
points in the $(\bar{x}, \bar{y})$ plane, corresponding to 4 values of $\theta$
for $r=0.15$ and 8 values of $\theta$ for $r=0.3$ and $r=0.6$. 
For the \bdksp\ mode, an additional set of 8 points with $r=0.9$ is added. 
200 toy MC experiments are generated at each point. The standard 
deviation and offset are then parametrized with polynomial functions 
in $\bar{x}$ and $\bar{y}$. 


The offset in $x$ and $y$ is large enough to affect the result of the
measurement (up to $0.02$ for the \bdk\ mode); it is larger in
the \bdskp\ and \bdksp\ modes, where the data samples are smaller. To
test the effect of sample size, toy MC experiments were generated
with event samples ten times larger than those of the \bdk\ 
data: the offset for this sample is less than $5\times 10^{-3}$. 
This is consistent with the maximum likelihood fit being unbiased
in the limit of large sample sizes. For each mode, the offset is
included in the PDF for the reconstructed parameters, so the
confidence intervals (and central values) for $\phi_3$, $r$, $\delta$ 
are unbiased. 

Different implementations of the frequentist technique can be used to 
obtain the values of the physical parameters. 
The two most widely used are central (so-called Neyman)
intervals, and the unified approach of Feldman and Cousins \cite{fc}.
The latter takes unphysical regions of the parameter
space into account. In this analysis, fitting with four parameters
$(x_{\pm},y_{\pm})$ in a problem defined by three parameters 
($r$, $\phi_3$, $\delta$) creates an unphysical region: 
all $(x_{\pm},y_{\pm})$ values where $r_+ \neq r_-$. 
For two of the three samples (\bdskp\ and \bdksp) 
the values of $r_+$ and $r_-$ differ significantly, {\it i.e.}\ 
the result for these modes is well into the unphysical region.
Central intervals would overestimate the significance of the
measurement in this case. Therefore, to obtain the values 
of the physical parameters $r$, $\phi_3$ and $\delta$
(we will refer to this set of parameters as a vector $\mu$)
given the measurement result $x_+$, $y_+$, $x_-$ and $y_-$ (or vector $z$)
we use the Feldman-Cousins approach. 


The confidence level $\alpha$ is calculated as
\begin{equation}
  \alpha(\mu)=\int\limits_{\mathcal{D}(\mu)}p(z|\mu)dz, 
  \label{fccl}
\end{equation}
where the integration domain $\mathcal{D}$ is given by likelihood ratio 
ordering:
\begin{equation}
  \frac{p(z|\mu)}{p(z|\mu_{\rm best}(z))}>\frac{p(z_0|\mu)}{p(z_0|\mu_{\rm best}(z_0))}. 
  \label{fcorder}
\end{equation}
Here, $p(z|\mu)$ is the normalized probability density to obtain the 
measurement result
$z$ for a given set of true parameters $\mu$. This PDF is obtained by toy MC
simulation. $\mu_{\rm best}(z)$ stands for the best true parameters for a given 
measurement $z$, {\it i.e.}\ $\mu$ such that $p(z|\mu)$ is maximized for the 
given $z$. $z_0$ is the result of the fit to experimental data. 

\begin{table}
  \caption{$CP$ fit results. The error intervals are statistical only. }
  \label{fit_res_table}
  \begin{tabular}{|l||c|c||c|c||c|c|} \hline
  Parameter & \multicolumn{2}{|c||}{\bdkp\ mode}
            & \multicolumn{2}{|c||}{\bdskp\ mode}
            & \multicolumn{2}{|c|}{\bdksp\ mode}\\
  \cline{2-7}
            & $1\sigma$ interval & $2\sigma$ interval & 
              $1\sigma$ interval & $2\sigma$ interval &
              $1\sigma$ interval & $2\sigma$ interval \\ \hline
  $\phi_3$  & $65.5^{\circ}\;^{+19.1^{\circ}}_{-19.9^{\circ}}$ 
            & $20.8^{\circ}<\phi_3<108.4^{\circ}$ 
            & $86.1^{\circ}\;^{+37.1^{\circ}}_{-93.1^{\circ}}$ 
            & - 
            & $10.8^{\circ}\;^{+22.7^{\circ}}_{-57.1^{\circ}}$ 
            & -
            \\
  $r$       & $0.165^{+0.056}_{-0.059}$
            & $0.048<r<0.281$ 
            & $0.207^{+0.127}_{-0.131}$
            & $r<0.510$ 
            & $0.549^{+0.231}_{-0.163}$
            & $0.242<r<1.120$ 
            \\
  $\delta$  & $147.5^{\circ}\;^{+18.7^{\circ}}_{-20.3^{\circ}}$
            & $100.3^{\circ}<\delta<187.7^{\circ}$ 
            & $334.7^{\circ}\;^{+36.8^{\circ}}_{-93.6^{\circ}}$
            & - 
            & $210.2^{\circ}\;^{+24.2^{\circ}}_{-57.3^{\circ}}$
            & - 
            \\
  \hline
  \end{tabular}
\end{table}

\begin{figure}[p]
  \epsfig{figure=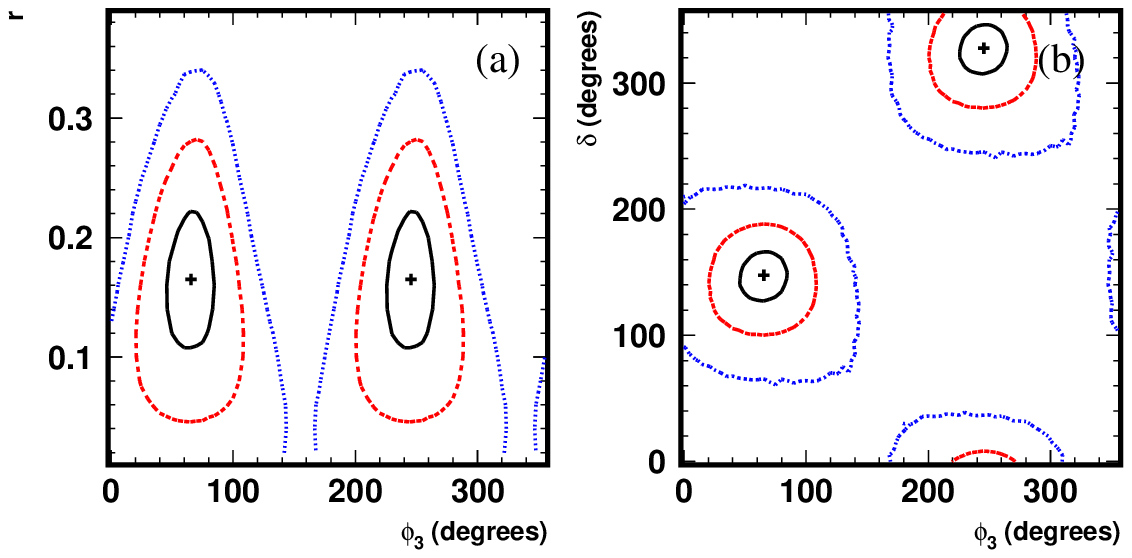,width=0.8\textwidth}
  \epsfig{figure=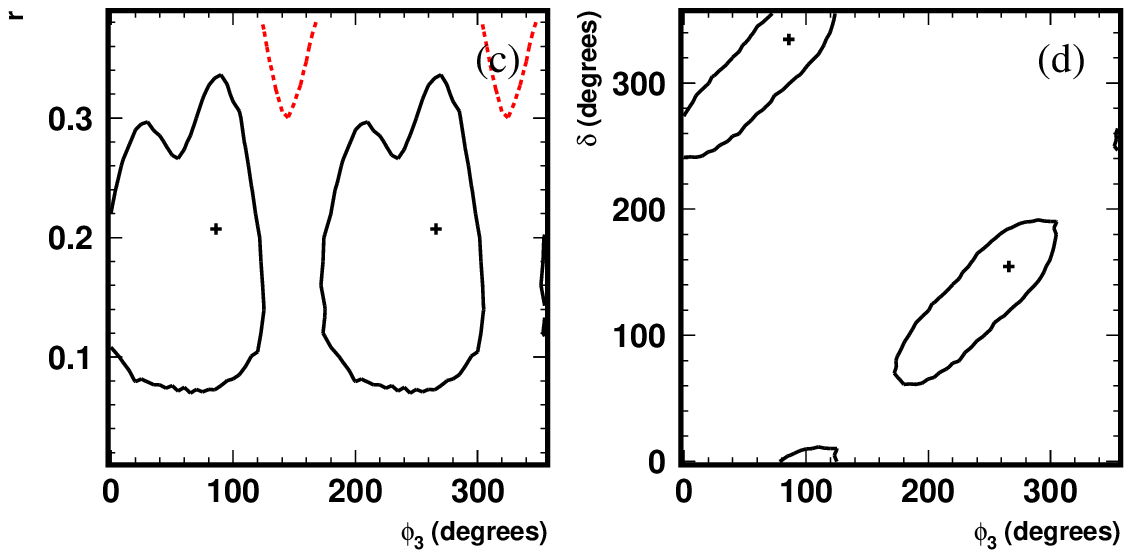,width=0.8\textwidth}
  \epsfig{figure=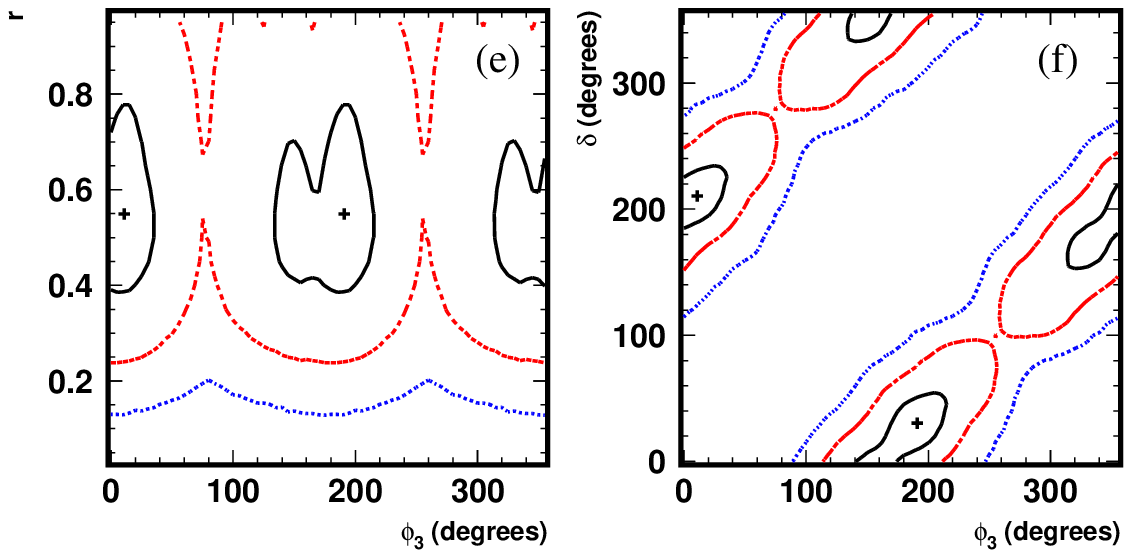,width=0.8\textwidth}
  \caption{Projections of confidence regions for the (a,b) \bdkp, 
           (c,d) \bdskp, and (e,f) \bdksp\ modes onto the 
           $(r, \phi_3)$ and $(\phi_3, \delta)$ planes. 
           Contours indicate integer 
	   multiples of the standard deviation. }
  \label{cont_fc}
\end{figure}

Fig~\ref{cont_fc} shows the projections of the three-dimensional 
confidence regions onto $(r, \phi_3)$ and $(\phi_3, \delta)$ planes. 
We show the 20\%, 74\% and 97\% confidence level regions, 
which correspond to one, two, and three standard deviations for a 
three-dimensional Gaussian distribution. The central values for the 
parameters $r$, $\phi_3$ and $\delta$ with their one and two standard 
deviation intervals are presented in Table~\ref{fit_res_table}. 
The confidence intervals in Table~\ref{fit_res_table} are calculated 
from projections of the three-dimensional confidence regions onto 
each of the parameters. Toy MC studies of the \bdk\ 
mode show that these intervals include the true value of $\phi_3$
slightly more often than the confidence levels imply (overcoverage), 
and are therefore conservative.

\subsection{Estimation of systematic errors}

Experimental systematic errors come from uncertainty in the knowledge of 
the functions used in the signal Dalitz plot fit. These include the Dalitz 
plot profiles of the backgrounds and the detection efficiency, the momentum 
resolution description, and, since the 
background fraction varies depending on the $\Delta E$ and $M_{\rm bc}$ parameters, 
the parametrizations of the $\Delta E-M_{\rm bc}$ shape of the signal and background. 

The systematic errors related to the Dalitz plot shape of the background are
studied by using different background descriptions in the signal fit. 
Instead of the baseline background shape, 
in which the continuum background is determined from the continuum-enriched
experimental sample, and the 
$b\bar{b}$ background from a MC sample, we perform fits with 
the background shape represented by only the continuum background 
parametrization, and the background shape extracted from the $M_{\rm bc}$
sideband ($5.2$ GeV/$c^2<M_{\rm bc}<5.26$ GeV/$c^2$). The maximum deviation of the 
parameters $x,y$ from those obtained with the ``standard" background 
parameterization was taken as a measure of the systematic error.

The distribution of the relative detection efficiency over the Dalitz plot is 
obtained from the MC simulation. To estimate the effect of imperfect 
simulation of the detector response, we also attempt to obtain the efficiency 
shape from the experimental data. We expect that the detection efficiency 
depends primarily on the momentum of the $D$ meson. We also assume that for
high momentum $D$ mesons, the efficiency does not vary significantly over 
phase space and is well determined in the simulation. Therefore, to 
obtain the relative efficiency distribution for, {\it e.g.}\ the \bdkp\ sample, 
we use the continuum \dsdpims\ sample with \dkpp\ (the one used for the determination
of the \dkpp\ amplitude) and compare Dalitz plots for high momentum 
$\overline{D}{}^0$ ($p_{D}>3.5$ GeV/$c$) and $\overline{D}{}^0$ in the 
momentum range corresponding to the \bdkp\ decay 
($2.1$ GeV/$c<p_{D}<2.4$ GeV/$c$). If the background is subtracted, 
the ratio of the event densities for the decays of $\overline{D}{}^0$ 
in different momentum ranges should equal the ratio of detection 
efficiencies. Once we assume that the efficiency for high momentum 
$\overline{D}{}^0$ is known, the efficiency 
distribution of $\overline{D}{}^0$ from \bdkp\ decay can be calculated. 
The difference of the parameters $x,y$ between fits using
MC-based and data-based efficiency shapes is taken as the
corresponding systematic error. 

The uncertainty due to momentum resolution is estimated by performing
a $\overline{D}{}^0$ model fit with a subsequent fit to $B$ decay data
without taking the momentum resolution into account. The bias of the 
$(x,y)$ parameters is negligible, therefore, this systematic uncertainty
is neglected. 

The uncertainty of the $\Delta E-M_{\rm bc}$ distribution parameterization
(which, in particular, leads to uncertainty in the signal-to-background
ratio) is taken into account by performing fits 
using $\Delta E-M_{\rm bc}$ shapes with each of the parameters varied by
one standard deviation. 

The $\Delta E-M_{\rm bc}$ shape and the total background fraction
are extracted from the signal $\Delta E-M_{\rm bc}$ fit. Both continuum 
and $b\bar{b}$ backgrounds are parametrized by a common $\Delta E-M_{\rm bc}$
density. However, in general, continuum and 
$b\bar{b}$ (and even each of the sub-components of these backgrounds) 
may be distributed differently, and the distribution may be correlated
with the Dalitz plot distribution. To estimate the contribution of this 
effect, we make additional fits to data with individual $\Delta E-M_{\rm bc}$
distributions for continuum and $b\bar{b}$ background components extracted
from generic MC samples. The corresponding bias is included in the background
systematic error. 

The results of the study of experimental systematic errors are summarized 
in Table~\ref{syst_table} for each mode. The uncertainties in 
the parameters $(x,y)$ are then used to obtain the systematic errors
for $r$, $\phi_3$ and $\delta$ in the combined measurement 
(see Section~\ref{section_combined}). In our previous 
analyses~\cite{belle_phi3_2,belle_phi3_3} the systematic error was 
dominated by a term due to the bias in the fit of $(\phi_3,r,\delta)$ 
to the \bdpip\ control sample (see Ref.~\cite{belle_phi3_2} Section V.B). 
In the present analysis, biases in the fit of $(x_{\pm},y_{\pm})$ to data 
are small, and are taken into account when forming confidence
intervals in $\phi_3$, $r$, and $\delta$ (section~\ref{section_stat}). 
Fitted values for the \bddspip\ control samples are consistent with
expectations (section~\ref{section_control}), with statistical precision 
comparable to the systematic errors listed in Table~\ref{syst_table}. 
No additional systematic error is assigned.

\begin{table}
  \caption{Systematic errors in $x,y$ variables. }
  \label{syst_table}
  \begin{tabular}{|l||c|c|c|c||c|c|c|c||c|c|c|c|}
    \hline
    Component & \multicolumn{4}{|c||}{\bdkp\ mode} 
              & \multicolumn{4}{|c||}{\bdskp\ mode} 
              & \multicolumn{4}{|c|}{\bdksp\ mode} \\ 
              \cline{2-13}
              & $\Delta x_-$ & $\Delta y_-$ & $\Delta x_+$ & $\Delta y_+$ 
              & $\Delta x_-$ & $\Delta y_-$ & $\Delta x_+$ & $\Delta y_+$ 
              & $\Delta x_-$ & $\Delta y_-$ & $\Delta x_+$ & $\Delta y_+$ \\
    \hline
    Background shape & 0.006 & 0.015 & 0.002 & 0.006 
                     & 0.011 & 0.023 & 0.002 & 0.004 
                     & 0.011 & 0.026 & 0.002 & 0.009 \\
    Efficiency shape & 0.011 & 0.004 & 0.017 & 0.001 
                     & 0.001 & 0.001 & 0.001 & 0.001 
                     & 0.024 & 0.023 & 0.002 & 0.009 \\
    $\Delta E-M_{\rm bc}$ stat. uncertainty 
                     & 0.002 & 0.005 & 0.002 & 0.004 
                     & 0.005 & 0.009 & 0.002 & 0.003 
                     & 0.006 & 0.013 & 0.003 & 0.003 \\
    $\Delta E-M_{\rm bc}$ difference for $q\bar{q}$ and $b\bar{b}$ 
                     & 0.004 & 0.002 & 0.000 & 0.005 
                     & 0.019 & 0.012 & 0.002 & 0.010 
                     & 0.010 & 0.027 & 0.004 & 0.001 \\
    \hline
    Total            & 0.013 & 0.016 & 0.017 & 0.009 
                     & 0.023 & 0.027 & 0.004 & 0.011 
                     & 0.029 & 0.046 & 0.006 & 0.013 \\
    \hline
  \end{tabular}
\end{table}

\subsection{Estimation of model uncertainty}

\label{section_model}

The model used for the \dkpp\ decay amplitude is one of the main sources of 
systematic error for our analysis. The amplitude is a result of 
the fit to the experimental Dalitz plot, however since the density of the 
plot is proportional to the absolute value squared of the decay amplitude, 
the phase 
of the complex amplitude is not directly measured. The phase variations across 
the Dalitz plot are therefore a function of model assumptions and 
their uncertainties may affect the $\tilde{D}$ Dalitz plot fit.

We use a MC simulation to estimate the effects of the model uncertainties. 
Event samples are generated according to the Dalitz distribution 
described by the amplitude given by Eq.~\ref{dalitz_density} 
with the resonance parameters extracted from 
our fit to continuum \dkpp\ data, but to fit this distribution
different models for $f(m^2_+, m^2_-)$ 
are used. We scan the phases $\phi_3$
and $\delta$ in their physical regions and take the maximum 
deviations of the fit parameters ($(\Delta r)_{\rm max}$, 
$(\Delta\phi_3)_{\rm max}$, and $(\Delta\delta)_{\rm max}$) 
as model uncertainty estimates. These studies use the value $r=0.15$, 
close to the one obtained in the dominant \bdkp\ mode. 

All the fit models are based on Breit-Wigner parameterizations 
of resonances as in our default model, but with differences in the
treatement of the broad components. 
By default, we use
Blatt-Weisskopf form factors for the $\overline{D}{}^0$ meson ($F_D$) and
intermediate resonances ($F_r$) and a $q^2$ dependence of the
resonance width $\Gamma$ (see~\cite{cleo_model}): to estimate the effect of the
corresponding theoretical uncertainties, we use a fit model
without form-factors, and with constant resonant widths $\Gamma_i$.
We also use a model containing only the largest 
doubly-Cabibbo-suppressed term $K^*(892)^-\pi^+$, the narrow 
resonances, $K^0_S f_0(980)$, and $K_0^*(1430)^+\pi^-$, with the 
remainder of the amplitude approximated by the flat non-resonant term.

Other models used are the model with all the resonances from 
Table~\ref{dkpp_table} excluding the $\sigma_1$ or $\sigma_2$ states, 
and the model used by CLEO \cite{dkpp_cleo}. 
The results of the study of model uncertainty are summarized in 
Table~\ref{model_table}. The maximum deviation of the parameters 
is taken as the uncertainty due to the \dkpp\ decay model. 
Each of the models listed in Table~\ref{model_table} has a reduced
$\chi^2$ significantly poorer than that of the default model. 
Therefore, our model uncertainty does not include an additional 
contribution due to the poor quality of the model fit, discussed 
in Section~\ref{section_d0_fit}. 


\begin{table}
\caption{Estimation of the \dkpp\ decay model uncertainty.}
\label{model_table}
\begin{tabular}{|l|c|c|c|} \hline
Fit model & $(\Delta r)_{\rm max}$ 
          & $(\Delta\phi_3)_{\rm max}$ ($^{\circ}$) 
          & $(\Delta\delta)_{\rm max}$ ($^{\circ}$) \\
\hline
$F_r=F_D=1$
    & 0.01 & 3.1 & 3.3 \\
$\Gamma(q^2)={\rm constant}$
    & 0.02 & 4.7 & 9.0 \\
$K^*(892)^+$, $\rho$, $\omega$, $K^*(892)^-$, $f_0(980)$, $K^*_0(1430)$, 
non-res.
    & 0.05 & 8.5 & 22.9 \\
No $\sigma_1$
    & 0.01 & 2.6 & 4.3 \\
No $\sigma_2$
    & 0.01 & 0.6 & 0.7 \\
CLEO model
    & 0.02 & 5.7 & 8.7 \\ \hline
\end{tabular}
\end{table}

The mode \bdksp\ has an additional uncertainty due to the possible 
presence of a nonresonant \bdkspnr\ component, which can also be treated 
as a model uncertainty. Since the nonresonant decay is described by 
the same set of diagrams as \bdksp, a similar $CP$ violating effect 
should be present that in general has values of $r$ and $\delta$
which differ from those for the resonant mode.
Thus, for a $\phi_3$ measurement without taking the \bdkspnr\ mode 
into account, its contribution can bias the fit parameters. 
To estimate the corresponding systematic uncertainty, 
we first measure the fraction of nonresonant decays within the \bdksp\ signal
region. To increase the sample size for this study, we include 
additional $\overline{D}{}^0$ decay modes: 
$K^+\pi^-$, $K^0_S\pi^+\pi^-$, $K^+\pi^-\pi^0$ and $K^+\pi^-\pi^+\pi^-$. 
Based on the $B^+$ yield in the $K^{*+}$ mass sidebands, 
and the observed shape of the invariant mass $M_{K^0_S\pi^+}$ of signal 
candidates, we find an upper limit of 6.3\% on the \bdkspnr\ fraction. 
We then perform a toy MC simulation with a 6.3\% nonresonant contribution 
added to determine the bias of the fit parameters. 
The fits are performed for various values of the $r$ and $\delta$ parameters 
of the nonresonant component and various values of the 
relative phase between \bdks\ and \bdksnr\ amplitudes. The maximum 
bias of the fit parameters is taken as the corresponding systematic error:
$\Delta r=0.084$, $\Delta\phi_3=8.3^{\circ}$, 
$\Delta\delta=49.3^{\circ}$. The $\phi_3$ bias is significantly smaller
than that for the strong phase $\delta$, since $\phi_3$ is obtained 
from a difference of the total phases for $B^+$ and $B^-$ decays, and
a part of the bias cancels in this case.

\subsection{Combined \boldmath{$\phi_3$} measurement}

\label{section_combined}

The three event samples, \bdkp, \bdskp, and \bdksp\ are combined 
in order to improve the sensitivity to $\phi_3$. 
The confidence levels for the combination of three modes are obtained 
using the same technique as for the single mode. 
In this case, the vector of physical parameters  
$\mu=(\phi_3, r_{DK}, \delta_{DK}, r_{D*K}, \delta_{D*K}, 
r_{DK*}, \delta_{DK*})$ and there are twelve measured parameters:
four parameters $(x_{\pm}, y_{\pm})$ for each of the three 
modes. The probability density of the measurement result $p(z|\mu)$ 
is the product of the probability densities for the individual modes. 

For each $\mu$, to obtain the confidence level $\alpha(\mu)$, 
the integral in Eq.~\ref{fccl} is performed over a 12-dimensional space. 
This requires extensive computation that makes it impractical to scan the whole 
range of physical parameters to obtain multidimensional confidence regions. 
However, it is still possible to calculate the confidence intervals for 
each of the individual parameters. 

To calculate the systematic errors for the combined measurement, 
we vary the measured parameters $x$ and $y$ within their systematic 
errors. Gaussian distributions are used for the variation, 
and the systematic biases are assumed to be uncorrelated. For each of the 
varied parameter sets, the central values of the physical parameters 
are calculated. The systematic error of the physical parameter
is then taken to be equal to the RMS of the resulting distribution. 

The error due to the uncertainty in the \dkpp\ amplitude is considered to 
be equal in all modes. The uncertainty due to the possible contribution of 
the nonresonant \bdkspnr\ amplitude to \bdksp\ decay is also included
in the model error, by varying the
parameters $(x_{\pm},y_{\pm})$ of the \bdksp\ sample according to the
biases $\Delta r$, $\Delta\phi_3$, and $\Delta\delta$ obtained above 
(Section~\ref{section_model}). We find a contribution of only $1.8^{\circ}$
to the combined $\phi_3$ measurement; this is added in
quadrature, yielding a total error of $8.7^{\circ}$.
For the $r$ and $\delta$ parameters of the \bdksp\ mode 
the \bdkspnr\ contribution dominates the model error. 

Confidence intervals for the combined measurement together 
with systematic and model errors are shown in Table~\ref{fc_comb_table}. 
The confidence intervals are statistical only and are calculated 
from projections of the seven-dimensional confidence regions onto 
each of the parameters. The statistical confidence level of $CP$ violation 
is 74\%. It is defined as the minimum 
confidence level $\alpha(\mu)$ for the $CP$-conserving set of 
physical parameters $\mu$ ({\it i.e.} the set with $\phi_3=0$). 
The two standard deviation interval (including model and systematic 
uncertainties) is calculated by adding the doubled systematic and 
model errors in quadrature to the two standard deviation statistical 
errors: we find $8^{\circ}<\phi_3<111^{\circ}$. 

We note that the two standard 
deviation interval of $\phi_3$ is more than twice as wide as the one standard 
deviation interval. The non-Gaussian errors in $\phi_3$ are related to the 
low significance of $r$; this effect should be reduced with larger 
data samples in the future. There is also a contribution from the disagreement 
between the \bdkp\ and \bdksp\ results, which leads to a second 
minimum in $\alpha(\mu)$ between the one and two standard deviation levels. 

\begin{table}
  \caption{Results of the combination of \bdkp, \bdskp, and \bdksp\ modes. }
  \label{fc_comb_table}
  \begin{tabular}{|l|c|c|c|c|c|} \hline
  Parameter & $1\sigma$ statistical interval & $2\sigma$ statistical interval & 
              Systematic error & Model uncertainty \\ \hline
  $\phi_3$  & $53.3^{\circ}\;^{+14.8^{\circ}}_{-17.7^{\circ}}$ 
            & $11.7^{\circ}<\phi_3<107.7^{\circ}$ 
            & $2.5^{\circ}$ & $8.7^{\circ}$ \\
  $r_{DK}$  & $0.159^{+0.054}_{-0.050}$
            & $0.048<r_{DK}<0.271$ & 0.012 & 0.049 \\
  $\delta_{DK}$  
            & $145.7^{\circ}\;^{+19.0^{\circ}}_{-19.7^{\circ}}$
            & $100.6^{\circ}<\delta_{DK}<185.9^{\circ}$ 
            & $3.0^{\circ}$ & $22.9^{\circ}$\\
  $r_{D^*K}$  & $0.175^{+0.108}_{-0.099}$ 
	      & $0<r_{D^*K}<0.407$ & 0.013 & 0.049 \\
  $\delta_{D^*K}$  & $302.0^{\circ}\;^{+33.8^{\circ}}_{-35.1^{\circ}}$
                   & - & $6.1^{\circ}$ & $22.9^{\circ}$\\
  $r_{DK^*}$  
            & $0.564^{+0.216}_{-0.155}$
            & $0.231<r_{DK^*}<1.106$ & 0.041 & 0.084\\
  $\delta_{DK^*}$  
            & $242.6^{\circ}\;^{+20.2^{\circ}}_{-23.2^{\circ}}$
            & $186.0^{\circ}<\delta_{DK^*}<300.2^{\circ}$ 
            & $2.5^{\circ}$ & $49.3^{\circ}$ \\ \hline
  \end{tabular}
\end{table}

\section{Conclusion}

We report the results of a measurement of the unitarity triangle angle 
$\phi_3$, using a method based on Dalitz plot analysis of
\dkpp\ decay in the process \bddsksp. The measurement of 
$\phi_3$ using this technique was performed based on a 357 fb$^{-1}$ 
data sample collected by the Belle detector. 
From the combination of \bdkp, \bdskp\ and \bdksp\ modes, we obtain the value
$\phi_3=53^{\circ}\;^{+15^{\circ}}_{-18^{\circ}}\pm 3^{\circ}\pm 9^{\circ}$; 
of the two possible solutions we choose the one with $0<\phi_3<180^{\circ}$.
The first error is statistical, the second is experimental systematics and
the third is model uncertainty. 
The two standard deviation interval (including model and systematic 
uncertainties) is $8^{\circ}<\phi_3<111^{\circ}$.
The statistical significance of $CP$ violation for the combined 
measurement is 74\%. 
The method allows us to obtain a value of the ratio of the two 
interfering $D$ decay amplitudes $r$, which can be used in other $\phi_3$ 
measurements. We obtain $r=0.159^{+0.054}_{-0.050}\pm 0.012\pm 0.049$
for the \bdkp\ mode, $r=0.175^{+0.108}_{-0.099}\pm 0.013\pm 0.049$ 
for the \bdskp\ mode and $r=0.564^{+0.216}_{-0.155}\pm 0.041\pm 0.084$ 
for the \bdksp\ mode. 

\section*{Acknowledgments}

We thank the KEKB group for the excellent operation of the
accelerator, the KEK cryogenics group for the efficient
operation of the solenoid, and the KEK computer group and
the National Institute of Informatics for valuable computing
and Super-SINET network support. We acknowledge support from
the Ministry of Education, Culture, Sports, Science, and
Technology of Japan and the Japan Society for the Promotion
of Science; the Australian Research Council and the
Australian Department of Education, Science and Training;
the National Science Foundation of China and the Knowledge 
Innovation Program of Chinese Academy of Sciencies under 
contract No.~10575109 and IHEP-U-503; the Department of Science and 
Technology of
India; the BK21 program of the Ministry of Education of
Korea, and the CHEP SRC program and Basic Research program 
(grant No. R01-2005-000-10089-0) of the Korea Science and
Engineering Foundation; the Polish State Committee for
Scientific Research under contract No.~2P03B 01324; the
Ministry of Science and Technology of the Russian
Federation; the Slovenian Research Agency;  
the Swiss National Science Foundation; the National Science Council and
the Ministry of Education of Taiwan; and the U.S.\
Department of Energy.

\appendix*
\section{Model uncertainty in $\bf(x,y)$ parameters}

Averaging the $\phi_3$ measurements between different methods and different experiments 
is most conveniently done using the $(x,y)$ parameters, which are better-behaved
than $(r,\phi_3,\delta)$. This requires errors 
(including model uncertainty) to be expressed in terms of the $(x,y)$ 
parameters, but our model uncertainty study was performed in $(r,\phi_3,\delta)$. 
To calculate the errors due to the \dkpp\ decay model uncertainty in $(x,y)$ we 
propagate the $(r,\phi_3,\delta)$ errors under the assumption that there is no 
correlation between them. This approach results in substantial off-diagonal terms. 
The correlation matrix for the vector 
$z=(x_{DK^-},y_{DK^-},x_{DK^+},y_{DK^+},x_{D^*K^-},y_{D^*K^-},x_{D^*K^+},y_{D^*K^+},
    x_{DK^{*-}},y_{DK^{*-}},x_{DK^{*+}},y_{DK^{*+}})$ is defined as
\[
  (\mathcal{C}_{D^0})_{ij}=\overline{\Delta z_i\Delta z_j}, 
\]
and is equal to
\begin{equation}
  \mathcal{C}_{D^0}=\left(
  \begin{array}{rrrrrrrrrrrr}
   459 &  10  & -104 & 332 &  -56 &   21 &   6  &  62  &  -12 &    68 &    64 &    30 \\
    10 & 241  & -232 & -64 &   -3 &    1 &   0  &   2  &    0 &     3 &     3 &     2 \\
  -104 & -232 &  264 & -68 &  -18 &    7 &   2  &  20  &   -3 &    22 &    20 &    10 \\
   332 &  -64 &  -68 & 436 &   55 &  -22 &   -4 &  -57 &    9 &   -66 &   -59 &   -30 \\
   -56 &  -3  &  -18 &  55 &  515 & -108 &  -54 &  400 &   10 &   -70 &   -64 &   -33 \\
    21 &   1  &    7 & -22 & -108 &  282 & -235 & -135 &   -4 &    27 &    25 &    13 \\
     6 &   0  &    2 &  -4 &  -54 & -235 &  242 &   26 &   -1 &     7 &     5 &     2 \\
    62 &   2  &   20 & -57 &  400 & -135 &   26 &  556 &  -13 &    76 &    69 &    33 \\
   -12 &   0  &   -3 &   9 &   10 &   -4 &   -1 &  -13 &  939 &    36 &  -328 &   878 \\
    68 &   3  &   22 & -66 &  -70 &   27 &    7 &   76 &   36 &   732 &  -556 &   -99 \\
    64 &   3  &   20 & -59 &  -64 &   25 &    5 &   69 & -328 &  -556 &   769 &   -86 \\
    30 &   2  &   10 & -30 &  -33 &   13 &    2 &   33 &  878 &   -99 &   -86 &   904 \\
  \end{array}
  \right)\times 10^{-5}.
\end{equation}

The uncertainty due to the non-resonant contribution to the 
\bdksp\ mode produces an additional contribution
\begin{equation}
  \mathcal{C}_{DK\pi}=\left(
  \begin{array}{rrrr}
 766 &  -371 &   113 &   828 \\
-371 &  2984 & -2605 & -1136 \\
 113 & -2605 &  2601 &   919 \\
 828 & -1136 &   919 &  1150 \\
  \end{array}
  \right)\times 10^{-5}.
\end{equation}
to the terms in $(x_{DK^{*-}},y_{DK^{*-}},x_{DK^{*+}},y_{DK^{*+}})$.


\end{document}